\documentclass[12pt]{article}
\usepackage{amsfonts,amssymb,amsmath}
\usepackage{epic,eepic}
\usepackage{theorem}
\usepackage{cite}
\newtheorem{definition}{Definition}[section]

\newtheorem{proposition}[definition]{Proposition}
\newtheorem{theorem}[definition]{Theorem}

\theorembodyfont{\rmfamily}

\numberwithin{equation}{section}

          \def\cB{{\cal B}}          
          \def\cE{{\cal E}}          
                    
                    \def\cL{{\cal L}}
                    
\def\cP{{\cal P}}                    
          \def\cT{{\cal T}}          \def\cU{{\cal U}}

\def\fB{{\mathfrak B}}

\newcommand{\be}{\begin{eqnarray}}
\newcommand{\ee}{\end{eqnarray}}
\newcommand{\CC}{{\mathbb C}}
\newcommand{\II}{{\mathbb I}}

\newcommand{\ZZ}{{\mathbb Z}}

\newcommand{\EE}{{\mathbb E}}

\newcommand{\finproof}{{\hfill \rule{5pt}{5pt}}}

\def\qmbox#1{\qquad\mbox{#1}\quad}

\def\tr{\mathop{\rm Tr}\nolimits}

\newcommand{\enne}{{\cal N}}
\newcommand{\emme}{{\cal M}}
\newcommand{\yy}{{\cal Y}}
\newcommand{\ff}{{\mathfrak f}}
\newcommand{\bs}[1]{{\boldsymbol{#1}}}
\newcommand{\wh}[1]{{\widehat{#1}}}
\newcommand{\wt}[1]{{\widetilde{#1}}}

\newcommand{\comaT}{{\cT^{*}}}
\newcommand{\comat}{T^{*}}

\begin{document}
\pagestyle{empty}

\null
\vfill
\begin{center}

{\Large \textsf{Analytical Bethe Ansatz
   for closed and open $gl(\enne)$-spin chains\\[1.2ex]
   in any representation}}

\vspace{10mm}

{\large D. Arnaudon$^a$, N. Cramp{\'e}$^{ab}$, A. Doikou$^a$, 
  L. Frappat$^{ac}$ and {\'E}. Ragoucy$^a$}
\footnote{daniel.arnaudon@lapp.in2p3.fr, nc501@york.ac.uk,
anastasia.doikou@lapp.in2p3.fr,\\
\null\qquad luc.frappat@lapp.in2p3.fr, eric.ragoucy@lapp.in2p3.fr}

\vspace{10mm}

\emph{$^a$ Laboratoire d'Annecy-le-Vieux de Physique Th{\'e}orique}

\emph{LAPTH, CNRS, UMR 5108, Universit{\'e} de Savoie}

\emph{B.P. 110, F-74941 Annecy-le-Vieux Cedex, France}

\vspace{7mm}

\emph{$^b$ University of York, Department of mathematics}

\emph{Heslington, York YO10 5DD, United Kingdom} 

\vspace{7mm}

\emph{$^c$ Member of Institut Universitaire de France}

\end{center}

\vfill
\vfill

\begin{abstract}
We present an ``algebraic treatment'' of the analytical Bethe Ansatz. 
For this purpose, we introduce abstract monodromy and transfer 
matrices which provide an algebraic framework for the analytical 
Bethe Ansatz. It allows us to deal with a generic $gl(\enne)$-spin chain 
possessing on each site an arbitrary $gl(\enne)$-representation. For 
open spin chains, we use the classification of the reflection matrices 
to treat all the diagonal boundary cases.\\
As a result, we obtain the Bethe equations in their full generality 
for closed and open spin chains. The classifications of finite 
dimensional irreducible representations for the Yangian (closed spin 
chains) and for the reflection algebras (open spin chains) are directly 
linked to the calculation of the transfer matrix eigenvalues.\\
As examples, we recover the usual closed and open spin chains, we 
treat the alternating spin chains and the closed spin chain with 
impurity.
\end{abstract}

\vfill
MSC: 81R50, 17B37 ---
PACS: 02.20.Uw, 03.65.Fd, 75.10.Pq
\vfill

\rightline{\texttt{math-ph/0411021}}
\rightline{LAPTH-1069/04}
\rightline{October 2004}

\baselineskip=16pt

\newpage

\pagestyle{plain}
\setcounter{page}{1}

\vspace*{1cm}
\section{Introduction}

The investigation of the integrable quantum spin chains was initiated by H.
Bethe in 1931 \cite{bet} where he studied the closed spin $1/2$ Heisenberg
chain \cite{heisen}. Since then, numerous generalisations of this spin
chain have been introduced: 
anisotropic XXZ spin chain \cite{faddeev, korepin, KoIzBo};
spin 1 chains \cite{ZAFA,MENERI};
alternating spin  chains \cite{dewo,abad2};
spin chains with higher spins \cite{Kul,Tak,ow,KoIzBo,KuSu,Bytsko,Tsuboi}; 
spin $1/2$ chains with spin $1$ impurities \cite{anjo,fuka,YWang}. 
Correlation functions of this type of spin chains have been computed
in e.g. [\citen{JMMN,JimWa,essler,MaiTer,kit,KMST,KSTS}].
The framework of integrable
open spin chains has been developed in \cite{Gau,cherednik,sklyanin,alc}.

There exist different motivations to study generalisations of the
integrable spin chains. First, they describe dynamics which can be
computed exactly of quantum mechanical models. Indeed, new models have been
investigated to describe theoretically crystalline material in order to
compare with the experimental data (e.g. for the crystals
$MnCu(S_2C_2O_2)_2(H_2O)_3$ see \cite{gv}, $(VO)_2P_2O_7$
\cite{jjgj,br,pbas,wbf,kmyu} or $Cu_2(OH)_2CO_3$ \cite{jlm}). 
Spin chains
allows one to treat some limits of other models as the Hubbard model
\cite{gut}, the quantum chromodynamics theory \cite{lip} or integrable
relativistic quantum field theories \cite{mw,aff,zaza,fsz}. Finally, recent
developments in AdS/CFT correspondence have also put spin chain models in
foreground of string theory \cite{miza,beis,Zarembo}.

The increasing number of applications urge us to seek for a complete
treatment of (closed and open) spin chain models. Different schemes exist
for dealing with these problems, most of them relying on the Bethe Ansatz
(coordinate, algebraic, analytical or thermodynamical for the main ones).
We present here a formulation of the analytical Bethe Ansatz for
$gl(\enne)$ (closed and open) spin chains whatever the representation at
each site the quantum spins belong to. In particular, we unify by this way
all the generalisations of the XXX model. 
\\

The main results of this paper are the following:
\begin{itemize}
\item[-]The determination of the Bethe equations for a closed spin
chain model where each quantum spin is represented in an arbitrary
representation of $gl(\enne)$ (called generic closed spin chain).
\item[-] 
The computation of Bethe equations for any open spin chain constructed
from an arbitrary diagonal reflection matrix and a
generic closed spin chain.  
\item[-]For each of the above mentioned models, the calculation of
the underlying symmetry and the integrability of the models. 
\end{itemize}

This paper consists of two main sections. The first one is devoted to the
study of closed spin chain, while the second one deals with the case of
boundaries. The structure of these two parts is similar. We recall first
the algebraic settings (Yangians or boundary algebras) for the monodromy
matrix. Then, we use the classification of the representations of these
algebras to compute a represented transfer matrix. Next, we use
generalisation of analytical Bethe Ansatz to obtain the Bethe equations.
Finally, we work out some examples.

\section{Closed spin chain\label{sect:cschain}}

\subsection{The $R$ matrix\label{sect:Rmatrix}}

We will consider the $gl(\enne)$ invariant $R$ matrices \cite{yang,baxter}
\begin{eqnarray}
  R_{ab}(\lambda) =\II_\enne \otimes \II_\enne -
\frac{\hbar\;\cP_{ab}}{\lambda}
\;,
\label{r}
\end{eqnarray}
where $\cP_{ab}$ is the permutation operator
\begin{equation}
  \label{eq:P12}
  \cP_{ab} = \sum_{i,j=1}^\enne E_{ij} \otimes E_{ji} 
\end{equation}
and $\hbar$ is the deformation parameter. It is usually set to 1 in
the context of quantum groups (Yangians), and to $-i$ when dealing with spin
chain models: here, we leave it free.
$E_{ij}$ are the
elementary matrices with 1 in position $(i,j)$ and 0 elsewhere. 
{}From the algebraic point of 
view, the value of $\hbar$ is irrelevant (provided non-vanishing). It 
is in general set to 1 when studying Yangians, while it is set to $-i$ 
in the spin chains context. Here, we leave it free.

This $R$ matrix satisfies the following properties \\[2mm] 
\textit{(i) Yang--Baxter equation} \cite{mac,yang,baxter,korepin,KoIzBo}
\begin{eqnarray}
  R_{ab}(\lambda_{a}-\lambda_{b})\ R_{ac}(\lambda_{a})\ R_{bc}(\lambda_{b})
  =R_{bc}(\lambda_{b})\ R_{ac}(\lambda_{a})\ 
  R_{ab}(\lambda_{a}-\lambda_{b})
  \label{YBE}
\end{eqnarray}
\textit{(ii) Unitarity}
\begin{eqnarray}
  R_{ab}(\lambda)\ R_{ba}(-\lambda) = \zeta(\lambda)\,
  \II_\enne \otimes \II_\enne\label{uni1}\;,
\end{eqnarray}
where $R_{ba}(\lambda) =\cP_{ab} R_{ab}(\lambda) \cP_{ab} =
R_{ab}^{t_{a}t_{b}}(\lambda) = R_{ab}(\lambda)$ and
\begin{eqnarray}
  \zeta(\lambda) = \left(1-\frac{\hbar}{\lambda}\right)
\left(1+\frac{\hbar}{\lambda}\right)\;.
\end{eqnarray}
It obeys $[A_{a} A_{b},\ R_{ab}(\lambda)] = 0$ for $A\in End(\CC^\enne)$. \\

The $R$ matrix can be interpreted physically as a scattering matrix
\cite{zamo, korepin, faddeev} describing the interaction between two
solitons (viewed in this framework as low level excited states in a
thermodynamical limit of a spin chain) that carry the fundamental
representation of $gl(\enne)$.

\subsection{Yangian $\yy(gl(\enne))$\label{sect:ygln}}

We present in this section some definitions and properties of the
Yangian \cite{Drinfeld}
associated to the Lie algebra $gl(\enne)$ that will be used in 
the following.

The Yangian $\yy(gl(\enne))$ is the complex associative unital algebra with
the generators $\{T_{ij}^{(n)}|1\leq i,j\leq \enne, n\in \ZZ_{\geq 0}\}$
 subject to the defining relations
\begin{eqnarray}
\label{relcom}
[T_{ij}^{(r+1)}\,,\,T_{kl}^{(s)}]-[T_{ij}^{(r)}\,,\,T_{kl}^{(s+1)}]
=T_{kj}^{(r)}\,T_{il}^{(s)}-T_{kj}^{(s)}\,T_{il}^{(r)}\;,
\end{eqnarray}
where $r,s\in \ZZ_{\geq 0}$ and $T_{ij}^{(0)}=\delta_{ij}$.\\
The $R$ matrix previously introduced allows us
to encode the Yangian defining relations in a simple equation, called 
FRT exchange relation \cite{FRT}
\begin{eqnarray}
\label{RTT}
R_{ab}(\lambda_a-\lambda_b)\;\cT_a(\lambda_a)\;\cT_b(\lambda_b)=
\cT_b(\lambda_b)\;\cT_a(\lambda_a)\;R_{ab}(\lambda_a-\lambda_b)\;,
\end{eqnarray}
where the generators are gathered in the following matrix (belonging 
to $End(\CC^\enne)\otimes \yy(gl(\enne))[[\lambda^{-1}]]$)
\begin{eqnarray}
\label{def:T}
\cT(\lambda)=\sum_{i,j=1}^\enne E_{ij}\otimes 
T_{ij}(\lambda)
=\sum_{i,j=1}^\enne E_{ij}\otimes 
\sum_{r \geq 0} \frac{\hbar^{r}}{\lambda^r}~T_{ij}^{(r)}=
\sum_{r \geq 0} \frac{\hbar^{r}}{\lambda^r}~\cT^{(r)}
\;.
\end{eqnarray}
Using the commutation relations (\ref{RTT}), it is easy to show that 
$\cT^{(1)}$ generates a $gl(\enne)$ algebra.\\
In order to construct representations of $\yy(gl(\enne))$, the
following algebra homomorphism from $\yy(gl(\enne))$ to $\cU(gl(\enne))$
(universal enveloping algebra of $gl(\enne)$)
will be used\footnote{To be compatible
with the pseudo-vacuum as usually defined in the study of spin chain
models, the convention used here for the homomorphism
differs from the one introduced in \cite{twmolev}. The link between
the two conventions is provided by the Yangian 
automorphism $T(\lambda)\longmapsto T^t(-\lambda)$, where $^t$ is the
usual transposition.}
\begin{eqnarray}
\label{eval}
T_{ij}(\lambda)\longmapsto \delta_{ij}-\frac{\hbar\;e_{ji}}{\lambda}\;,
\end{eqnarray}
where $\{e_{ij}\}$ is a basis of the Lie algebra $gl(\enne)$.
The Yangian of $gl(\enne)$ is a Hopf algebra with the coproduct given by 
\begin{eqnarray}
\label{coproduct}
\Delta : \yy(gl(\enne))& \longrightarrow& \yy(gl(\enne))\otimes
\yy(gl(\enne))\nonumber\\
T_{ij}(\lambda)&\longmapsto& \sum_{k=1}^\enne T_{ik}(\lambda)\otimes
T_{kj}(\lambda)\;.
\end{eqnarray}
The coproduct is the cornerstone  to deal with the tensor product of
representations. We define also by recursion
$\Delta^{(n)}=(\Delta\otimes id^{\;\otimes\; n-2})\Delta^{(n-1)}$
for $n>2$ and $\Delta^{(2)}=\Delta$.\\

The quantum determinant $qdet\,\cT(\lambda)$ is a formal series in
$\lambda^{-1}$ with coefficients in $\yy(gl(\enne))$ defined
as follows
\begin{eqnarray}
\label{qdet}
qdet\,\cT(\lambda)=
\sum_{\sigma\in \mathfrak{S}_\enne}sgn(\sigma)~
T_{1,\sigma(1)}(\lambda-\hbar\enne+\hbar)
\cdots T_{\enne,\sigma(\enne)}(\lambda)\;,
\end{eqnarray}
where $\mathfrak{S}_\enne$ is the permutation group of $\enne$
indices. A well-known result (see
e.g. \cite{molev}) establishes that the coefficients of $qdet\;\cT(\lambda)$
are algebraically independent and generate the centre of
$\yy(gl(\enne))$.
It is important for the following to realise that the
quantum determinant represented in any finite-dimensional irreducible
representation will be proportional to the identity matrix. 

There exists an equivalent definition of the quantum determinant which
will be used in the following as well. Let
$A_m$ be the antisymmetriser operator in $(\CC^\enne)^{\otimes m}$,
i.e.
\begin{equation}
\label{antisym}
A_m(e_{i_1}\otimes\cdots\otimes e_{i_m})=\frac{1}{m!}
\sum_{\sigma\in\mathfrak{S}_m}
sgn(\sigma)~e_{i_{\sigma(1)}}\otimes\cdots\otimes e_{i_{\sigma(m)}}\;,
\end{equation}
where $\{e_i|1\le i \le \enne\}$ is the canonical basis of $\CC^\enne$
and $1\le i_1,\dots, i_m \le \enne$. The antisymmetriser
is a  projector in
$(\CC^\enne)^{\otimes m}$. It has the remarkable property:
\begin{proposition}\textbf{\cite{MNO}}
\label{comAT} The following identities hold
\begin{eqnarray}
A_m\;\cT_1(\lambda)\cdots
\cT_m(\lambda-m\hbar+\hbar)A_m
&=&\cT_m(\lambda-m\hbar+\hbar)\cdots
\cT_1(\lambda)\;A_m\\
&=&A_m\;\cT_1(\lambda)\cdots \cT_m(\lambda-m\hbar+\hbar)\;.
\end{eqnarray}
\end{proposition}
When $m=\enne$, the antisymmetriser
becomes a one-dimensional projector and one can show \cite{MNO}
\begin{equation}
\label{qdet2} 
qdet\;\cT(\lambda)\;A_\enne=\cT_\enne(\lambda-\hbar\enne+\hbar)
\cdots \cT_1(\lambda)\;A_\enne\;.
\end{equation}
The relation (\ref{qdet2}) can be used as an equivalent
definition of the quantum determinant.\\
To study spin chains, we will use the following automorphisms of
$\yy(gl(\enne))$ \\
{\it (i) Inversion}
\begin{eqnarray}
inv:\ \cT(\lambda)\longmapsto \cT^{-1}(-\lambda)
\end{eqnarray}
{\it (ii) Shift}
\begin{eqnarray}
 s_{a}:\ \cT(\lambda)\longmapsto \cT(\lambda+a)\,,\ a\in\CC\label{shift}
\;.
\end{eqnarray}
One can compute the elements of $\cT^{-1}(\lambda)$ in terms of
$\cT(\lambda)$ using the following formula
\begin{eqnarray}
\label{comp:inv}
\cT^{-1}(\lambda-\hbar\enne+\hbar)=\big(qdet\;\cT(\lambda)\big)^{-1}\;
\comaT(\lambda)\;,
\end{eqnarray}
where $\comaT(\lambda)$ is the quantum comatrix, i.e. its entries
$\comat_{ij}(\lambda)$ are $(-1)^{i+j}$ times the quantum determinants
of the submatrices of 
$\cT(\lambda)$ obtained by removing the $i^{th}$ column and $j^{th}$ row.

\subsection{Algebraic transfer matrix}

In the following, in order to construct spin chains, it will be
necessary to deal with the tensor product of $\ell$ copies of the
Yangian. For $1\leq i \leq \ell$, we denote by $\cL_{ai}(\lambda)\in End
(\CC^\enne) \otimes \yy(gl(\enne)) $ one copy of the Yangian which
acts non trivially on the $i^{th}$ space only. The space $a$, 
always isomorphic to $End(\CC^\enne)$ in the 
present paper, is called
 auxiliary space whereas the space $i$ is called  quantum
space. Obviously, $\cL_{ai}(\lambda)$ satisfies the defining relations
of the Yangian
\begin{eqnarray}
\label{Rll}
R_{ab}(\lambda_a-\lambda_b)\;\cL_{ai}(\lambda_a)\;\cL_{bi}(\lambda_b)=
\cL_{bi}(\lambda_b)\;\cL_{ai}(\lambda_a)\;R_{ab}(\lambda_a-\lambda_b)\;.
\end{eqnarray}
Let us stress that the matrix $\cL_{ai}(\lambda)$ is local, i.e. it 
contains only
the $i^{th}$ copy of the Yangian. On the contrary, thanks to the
coproduct, one constructs a non-local algebraic object, the monodromy
matrix
\begin{align}
\label{mono}
\cT_a(\lambda)&=\Delta^{(\ell)}(\cL(\lambda))=
\cL_{a1}(\lambda)\;\cL_{a2}(\lambda)\;\dots\;\cL_{a\ell}(\lambda)\in End
(\CC^\enne) \otimes \left(\yy(gl(\enne))\right)^{\otimes \ell} \;.
\end{align}
Let us remark that the quantum spaces are omitted in the LHS of
(\ref{mono}), as usual in the notation of the monodromy matrix. The
entries of the monodromy matrix $\cT_a(\lambda)$ are given by
\begin{align}
\label{mono2}
T_{ij}(\lambda)&=\sum_{k_1,\dots,\,k_{\ell-1}=1}^\enne
L_{ik_1}(\lambda)\;\otimes\;L_{k_1k_2}(\lambda)\;\otimes\;
\dots\;\otimes\;L_{k_{\ell-1}j}(\lambda)\;.
\end{align}
Since the coproduct is a morphism, $\cT_a(\lambda)$ also satisfies
the defining relations of the Yangian
\begin{eqnarray}
\label{Rtt}
R_{ab}(\lambda_a-\lambda_b)\;\cT_{a}(\lambda_a)\;\cT_{b}(\lambda_b)=
\cT_{b}(\lambda_b)\;\cT_{a}(\lambda_a)\;R_{ab}(\lambda_a-\lambda_b)\;.
\end{eqnarray}
Now, we can introduce the main object for the study of spin
chains, i.e. the transfer matrix 
\begin{eqnarray}
\label{transf}
t(\lambda)
=tr_{a}\left(\cT_{a}(\lambda)\right)
=\sum_{i=1}^\enne T_{ii}(\lambda)\;.
\end{eqnarray}
Equation (\ref{Rtt}) immediately implies
\begin{eqnarray}
\label{comt}
[\;t(\lambda)\;,\;t(\mu)\;]=0
\end{eqnarray}
which will guarantee the integrability of the models (see section
\ref{sect:ham}).

Let us remark that, at that point, the monodromy and transfer 
matrices are algebraic objects (in $\left(\yy(gl(\enne))\right)^{\otimes 
\ell}$), and, as such, play the r{\^o}le of generating functions for the 
construction of monodromy and transfer  matrices as they are usually 
introduced in spin chain models. The latter will be constructed from 
the former using representations of the Yangian, as it will be done 
below.

\subsection{Symmetry}

The algebraic structure defined above is sufficient to determine the symmetry
of the transfer matrix. Indeed, we have:
\begin{proposition}
  The $gl(\enne)$ algebra is a symmetry of $t(\lambda)$. Its
  generators are expressed in terms of the local $gl(\enne)$ generators as
  \begin{eqnarray}
    T^{(1)}_{ij}= L^{(1)}_{ij}\;\otimes\;1^{\otimes \ell-1}
    +1\;\otimes\;L^{(1)}_{ij}\;\otimes\;1^{\otimes \ell-2} +\dots+
    1^{\otimes \ell-1}\otimes\;L^{(1)}_{ij}~\;.\label{sym}
  \end{eqnarray}
  \label{prop:symt}
\end{proposition}    
\textbf{Proof:} Taking the trace in space $a$ of the exchange relations
(\ref{Rtt}), we obtain
\begin{eqnarray}
\label{trace:Rtt}
(\lambda_a-\lambda_b)~[\,t(\lambda_a)\;,\;\cT(\lambda_b)\,]=
\hbar\;[\,\cT(\lambda_a)\;,\;\cT(\lambda_b)\,]\;.
\end{eqnarray}
Then, the $\lambda_b$ free term reads
\begin{eqnarray}
\label{symt}
\left[\,t(\lambda_a)\;,\;\cT^{(1)}\,\right]=0\;, 
\end{eqnarray}
which proves that $\cT^{(1)}=\sum_{i,j=1}^\enne 
E_{ij}\otimes T^{(1)}_{ij}$ is a symmetry of the transfer matrix.
These generators generates the $gl(\enne)$ Lie algebra.
\finproof\\
Thus, anticipating the spin chain interpretation, we can deduce that 
\underline{all} the integrable models constructed in the usual way 
from $t(\lambda)$ (such as the ones
presented in section \ref{sect:ex}) possess a 
$gl(\enne)$ symmetry. In other words, the $gl(\enne)$ symmetry is 
valid whatever the Yangian representations are. Depending on the 
model considered (i.e. the choice of representations), we will get the
 expression of the symmetry generators  
 by evaluating the relation (\ref{sym}) in the 
representations under consideration.

\subsection{Representations \label{reps}}

As already mentioned, spin chain models will be obtained through the 
evaluation of the algebraic monodromy and transfer matrices in 
Yangian representations. We thus present here some basic results on 
the classification of finite-dimensional irreducible representations 
of $\yy(gl(\enne))$.

\subsubsection{Evaluation representations}

Keeping in mind the forthcoming spin chains interpretation, we choose for
each local $\yy(gl(\enne))$ algebra an irreducible finite-dimensional
evaluation representation. 

We start with a finite-dimensional irreducible representation of
$gl(\enne)$, $M({\bs \alpha})$, with highest weight ${\bs \alpha} =
(\alpha_1,\dots,\alpha_\enne)$ and associated to the highest weight vector
$v$. This highest weight vector obeys
\begin{eqnarray}
&&e_{kj}\;v=0 \qmbox{,} 1\leq k<j \leq \enne  \\
&&e_{kk}\;v=\alpha_k\; v \qmbox{,}  1\leq k \leq \enne \;,
\end{eqnarray}
where $\alpha_1,\dots,\alpha_\enne$ are integers with
$\alpha_{k+1}\leq \alpha_k$. Indeed the constraints on the parameters
$\alpha_k$ are criteria so that the representation be
finite-dimensional and irreducible. Similar criteria will be given in
Theorem \ref{the:irr} for the Yangian.

The evaluation representation $M_\lambda({\bs \alpha})$ of 
$\yy(gl(\enne))$ is built from $M({\bs \alpha})$ and
follows from the homomorphism (\ref{eval}), according to
\begin{eqnarray}
&&L_{jk}(\lambda)\;v=0 \qmbox{,} 1\leq k<j \leq \enne  \\
&&L_{kk}(\lambda)\;v=\left(1-\frac{\hbar\;\alpha_k}{\lambda}\right)\; v 
\qmbox{,}  1\leq k \leq \enne \;.
\end{eqnarray}
It is important for the following to remark that the previous
relations imply that the entries of the
matrix $\lambda \cL(\lambda)$ are analytical.

The representation $M_\lambda((1,0,\dots,0))$, associated to the $gl(\enne)$
fundamental representation, of $\cL(\lambda)$ 
provides the $R$ matrix (\ref{r}).\\
Let us remark that $M_{\lambda+a}({\bs \alpha})~~(a\in \CC)$ defines
also a representation of the Yangian, which is isomorphic to 
$M_{\lambda}({\bs \alpha})$, according to the shift automorphism 
(\ref{shift}).

\subsubsection{Representations of the monodromy matrix \label{decadix}}

The evaluation representations of $\cL(\lambda)$ allow us to build a
representation of the monodromy matrix. Indeed, evaluating each of the
local $\cL_{a,n}(\lambda)$ in a representation
$M_{\lambda+a_{n}}(\bs{\alpha^n})$ for $1\leq n\leq\ell$, the tensor
product built on
\begin{eqnarray}
\label{tensorp}
M_{\lambda+a_{1}}(\bs{\alpha^1}) \otimes \dots \otimes
M_{\lambda+a_{\ell}}(\bs{\alpha^\ell})
\end{eqnarray}
provides, via (\ref{mono2}), a finite-dimensional representation  for 
$\cT(\lambda)$.\\
Denoting by $v^n$ the highest weight vector associated to
$\bs{\alpha^n}=(\alpha^n_1,\dots,\alpha^n_\enne)$, the vector 
\begin{equation}
    v^+=v^1 \otimes \dots \otimes v^\ell
\label{v+}
\end{equation}
is the highest weight vector of the representation (\ref{tensorp}) i.e.
\begin{eqnarray}
  &&T_{jk}(\lambda)\;v^+=0 \qmbox{,} 1\leq k<j \leq \enne  \\
  &&T_{kk}(\lambda)\;v^+=\prod_{n=1}^\ell
  \left(1-\frac{\hbar\;\alpha^n_k}{\lambda+a_{n}}\right)\;v^+ \qmbox{,}
  1\leq k \leq \enne \;.
\label{HWmono2}
\end{eqnarray}
We will be interested only in the irreducible finite-dimensional
representations of the monodromy matrix. When the representation is
reducible, the Bethe Ansatz does not give all the eigenvalues of the
transfer matrix.

There exists a necessary and sufficient criteria for a tensor product of
Yangian representations to be irreducible. It uses the following definition
\begin{definition}
  Let $X$ and $Y$ two disjoint finite subsets of $\ZZ$. $X$ and $Y$ are
  crossing if there exists $x_1,x_2\in X$ and $y_1,y_2\in Y$ such that
  \begin{eqnarray}
    x_1<y_1<x_2<y_2 \qmbox{or} y_1<x_1<y_2<x_2\;.
  \end{eqnarray}
  Otherwise $X$ and $Y$ are non-crossing.
\end{definition}
We associate to each highest weight $\bs{\alpha}$ the following subset
of $\ZZ$
\begin{eqnarray}
  X_\bs{\alpha}=\{\alpha_1,\alpha_2-1,\dots,\alpha_\enne-\enne+1\}~.
\end{eqnarray}
The theorem giving the criteria to obtain irreducible representations
states:
\newpage
\begin{theorem} \textbf{\cite{molirr}}
  \label{the:irr}
  \begin{enumerate}
    \item The tensor product
    $M_{\lambda+a_{1}}(\bs{\alpha^1})\otimes\dots\otimes
    M_{\lambda+a_{\ell}}(\bs{\alpha^\ell})$ is irreducible if and only if
    the tensor product
    $$M_{\lambda}(\bs{\alpha^1}-\frac{a_{1}}{\hbar}\bs{1})\otimes\dots\otimes
    M_{\lambda}(\bs{\alpha^\ell}-\frac{a_{\ell}}{\hbar}\bs{1})\,$$ is
    irreducible, with $\bs{1}=(1,\ldots,1)$.
    \item
    The tensor product $M_\lambda(\bs{\alpha^1})\otimes\dots\otimes
    M_\lambda(\bs{\alpha^\ell}) $ is irreducible if and only if all the
    tensor products $M_\lambda(\bs{\alpha^p})\otimes
    M_\lambda(\bs{\alpha^q})$ with $p<q$ are irreducible. 
    \item 
    The tensor product $M_\lambda(\bs{\alpha}) \otimes
    M_\lambda(\boldsymbol{\beta})$ is irreducible if and only if the sets
    $X_\bs{\alpha} \backslash X_\bs{\beta}$ and $X_\bs{\beta} \backslash
    X_\bs{\alpha}$ are non-crossing.
  \end{enumerate}
\end{theorem} 
Note that if 
$\bs{\alpha^{1}}=\bs{\alpha^{2}}=\ldots=\bs{\alpha^{\ell}}\equiv\bs{\alpha}$
the tensor product $M_\lambda(\bs{\alpha})^{\otimes\ell}$ is 
irreducible. This special case is the one generally used for spin 
chains models (see examples below).

\subsection{Analytical Bethe Ansatz}

We now use the above mathematical framework to study general closed spin
chains. We will be able to construct and study a spin chain with arbitrary
(and not necessarily identical) representations of $gl(\enne)$ on each site
of the chains. Put in other words, the algebraic set-up given above allows
us to treat simultaneously all the possible spin chain models built in this
way. In particular, we will obtain the Bethe equations for all these
models.

\subsubsection{Hamiltonian of the model \label{sect:ham}}

{}From now on, we use as local and monodromy matrices the following
elements:
\begin{eqnarray}
  \label{normal}
  \wh\cL_{a,n}(\lambda)=(\lambda+a_{n})\cL_{a,n}(\lambda) \qmbox{and}
  \wh\cT(\lambda)= \prod_{n=1}^\ell(\lambda+a_{n})\,\cT(\lambda)\,.
\end{eqnarray}
The normalisation of the monodromy matrix (\ref{normal}) ensures its
analyticity. Such a condition is crucial for the analytical
Bethe Ansatz method. The transfer matrix will be accordingly normalised: $\wh
t(\lambda)=tr_{a}\wh\cT_{a}(\lambda)$.

The properly normalised transfer matrix is a monic polynomial in $\lambda$
of degree $\ell$: $\displaystyle\wh t(\lambda) =
\lambda^\ell+\sum_{n=0}^{\ell-1} H_{n}\lambda^n$. The $\ell$ generalised
Hamiltonians $H_{n}$ are in involution (see equation (\ref{comt})) and
algebraically
independent (proved by looking at the number of involved sites in each
$H_{n}$). The Hamiltonian of the spin chain model under consideration will
be constructed as a polynomial in the generalised Hamiltonians $H_{n}$ and
will be then integrable.

Usually, in the spin chain context, we deal with Hamiltonian describing a
local interaction, i.e. an interaction between nearest neighbour.
Unfortunately, at this stage, there is no explicit formula to compute this
type of Hamiltonian from the transfer matrix. Note however that when all
the quantum spaces correspond to the same representation, an approach using
the fusion of auxiliary spaces can be applied \cite{Kul}. Explicit forms of
Hamiltonians will be also given for various models in section
\ref{sect:ex}.

\subsubsection{Highest weight vector\;/\; Pseudo-vacuum}

We now compute the eigenvalues of the transfer matrix $\wh
t(\lambda)$. As a  by-product, they will provide the Hamiltonian
eigenvalues. The procedure consists in three steps.

The first step consists in finding a particular eigenvector (so-called
pseudo-vacuum) of the transfer matrix and in computing the corresponding
eigenvalue. We get
\begin{eqnarray}
  &&\wh T_{jk}(\lambda)\;v^+=0 \qmbox{,} 1\leq k<j \leq \enne  \\
  &&\wh T_{kk}(\lambda)\;v^+=\prod_{n=1}^\ell
  \left(\lambda+a_{n}-\hbar\;\alpha^n_k\right)\;v^+ \qmbox{,} 
  1\leq k \leq \enne \;,
\end{eqnarray}
where $v_{+}$ is given in (\ref{v+}).\\
In the following, we use the following notation, for $1\leq k \leq
\enne$ 
\begin{eqnarray}
P_k(\lambda)=\prod_{n=1}^\ell
\left(\lambda+a_{n}-\hbar\;\alpha^n_k\right)\;.\label{drinP}
\end{eqnarray}
These polynomials, called Drinfel'd polynomials, are usually introduced
 to classify the representations of Yangians.\\

The highest weight vector (\ref{v+}) is obviously an eigenvector of the
transfer matrix. Indeed, one gets
\begin{eqnarray}
\wh t(\lambda)\;v^+=\sum_{k=1}^\enne ~\wh T_{kk}(\lambda)\;v^+
=\Lambda^0(\lambda)\;v^+
\end{eqnarray}
with
\begin{eqnarray}
\Lambda^0(\lambda)=\sum_{k=1}^\enne ~
P_k(\lambda)\;.
\end{eqnarray}
Note that $\Lambda^0(\lambda)$ is analytical. In the context of the spin
chains, the highest weight vector $v^+$ is called the pseudo-vacuum. The
second step consists in the Ansatz itself which provides all the
eigenvalues of $\wh t(u)$ from $\Lambda^0(\lambda)$.

\subsubsection{Dressing functions}

We make the following assumption for the structure of all the
eigenvalues of $\wh t(u)$ 
\begin{eqnarray}
\Lambda(\lambda)=\sum_{k=1}^\enne ~
P_k(\lambda)\;
D_k(\lambda)\;,\label{ClosedEigen}
\end{eqnarray}
where $D_k(\lambda)$, the so-called dressing functions, have to be
determined. At that point, the irreducibility of the 
representation is a necessary criteria for the completeness of the 
spectrum obtained by dressing.\\
{}From the asymptotic behaviour ($\lambda\rightarrow +\infty$) of 
$\wh t(\lambda)$, we
deduce that, for $1\leq k \leq \enne$ 
\begin{eqnarray}
D_k(\lambda)\xrightarrow[\lambda\rightarrow +\infty]{}1\;.
\end{eqnarray}
We suppose that the dressing functions are rational functions of the 
form
\begin{eqnarray}
\label{dress:funct}
D_k(\lambda)=\prod_{n=1}^{M^{(k-1)}}
\frac{\lambda+u_n^{(k-1)}}{\lambda-\lambda_n^{(k-1)}-\frac{\hbar\;(k-1)}{2}}
\prod_{n=1}^{M^{(k)}}
\frac{\lambda+v_n^{(k)}}{\lambda-\lambda_n^{(k)}-\frac{\hbar\;k}{2}}\;,
\end{eqnarray}
where $M^{(0)}=M^{(\enne)}=0$.\\ 
\textbf{Remarks}
\begin{enumerate}
  \item 
  The relation between $D_{k}(\lambda)$ and $D_{k+1}(\lambda)$ poles is the
  basic ingredient of the analytical Bethe Ansatz. This pole structure is
  the simplest one which ensures the analyticity of the eigenvalues. 
  \item
  We introduced shifts in the denominators for later convenience. 
  \item 
  The Lie algebra $gl(\enne)$ being an invariance of the transfer matrix
  (see proposition \ref{prop:symt}), the transfer matrix eigenvectors are
  indeed eigenvectors of the $gl(\enne)$ Cartan generators. The numbers
  $M^{(k)}$ ($1\leq k \leq \enne-1$) are deduced from the action of these
  Cartan generators on the eigenvector of eigenvalue $\Lambda(\lambda)$.
\end{enumerate}
We now tackle the third step, which consists in finding constraints to
determine $u_{n}^{(k)}$ and $v_{n}^{(k)}$ in terms of $\lambda_{n}^{(n)}$.

\subsubsection{Fusion procedure}

We shall use the fusion introduced previously in \cite{lepetit,selene} to
obtain constraints on the dressing functions.

Let $A_\enne$ be the antisymmetriser defined by the relation
(\ref{antisym}) which acts on auxiliary spaces $a_1,\dots,a_\enne$. Then,
from the following relation
\begin{eqnarray}
\wh \cT_{a_\enne}(\lambda-\hbar\enne+\hbar)
\cdots \wh \cT_{a_1}(\lambda)
~=~qdet\;\wh \cT(\lambda)~A_\enne
+
\wh \cT_{a_\enne}(\lambda-\hbar\enne+\hbar)
\cdots \wh \cT_{a_1}(\lambda)\;(1-A_\enne)\;,
\end{eqnarray}
we deduce, by taking the trace in the spaces $a_1,\dots,a_\enne$, that
\begin{eqnarray}
  \label{relfusion}
  \wh t(\lambda-\hbar\enne+\hbar)~\wh t(\lambda-\hbar\enne+2\hbar) ~\dots~
  \wh t(\lambda) =qdet\;\wh \cT(\lambda)+\wh t_\ff(\lambda)\;,
\end{eqnarray}
where $\wh t_\ff(\lambda)=tr_{a_1\dots a_\enne} \wh
\cT_{a\enne}(\lambda-\hbar\enne+\hbar) \cdots \wh
\cT_{a_1}(\lambda)(1-A_\enne)$ is the so-called fused transfer matrix.\\
We can compute the value of the quantum determinant using (\ref{qdet}) and
the properties of the highest weight. Indeed,
\begin{eqnarray}
  qdet\;\wh \cT(\lambda)~v^+&=&\sum_{\sigma\in
  \mathfrak{S}_\enne}sgn(\sigma)~ \wh
  T_{1,\sigma(1)}(\lambda-\hbar\enne+\hbar) \cdots \wh
  T_{\enne,\sigma(\enne)}(\lambda)~v^+\\
  &=&\prod_{k=1}^\enne~ P_k(\lambda-\hbar\enne+\hbar k)~v^+\;.
  \label{qdetv+}
\end{eqnarray}
The quantum determinant being central, the above relation implies that
\begin{eqnarray}
qdet\;\wh \cT(\lambda)=\prod_{k=1}^\enne~ P_k(\lambda-\hbar\enne+\hbar
k)\,. \label{eq:qdet}
\end{eqnarray}
Then, acting with any eigenvector $v$ with eigenvalue
$\Lambda(\lambda)$ on relation (\ref{relfusion}), one obtains
\begin{eqnarray}
  \label{fuslam}
  \Lambda(\lambda-\hbar\enne+\hbar)\dots \Lambda(\lambda)
  =\prod_{k=1}^\enne~P_k(\lambda-\hbar\enne+\hbar k)
  +\Lambda_\ff(\lambda)\;,
\end{eqnarray}
where $\Lambda_\ff(\lambda)\;v=\wh t_\ff(\lambda)\;v$. Let us remark that
this relation shows that $v$ is also an eigenvector of $\wh
t_\ff(\lambda)$, in accordance with the commutator
\begin{eqnarray}
  [\;\wh t_\ff(\lambda)\;,\;\wh t(\mu)\;]=0\;.
\end{eqnarray}
Finally, picking the term proportional to $\prod_{k=1}^\enne~
P_k(\lambda-\hbar\enne+\hbar k)$ in the relation (\ref{fuslam}), we deduce
a constraint between the dressing functions, namely
\begin{eqnarray}
  D_1(\lambda-\hbar\enne+\hbar)\dots D_\enne(\lambda)=1\;.
\end{eqnarray}
This constraint allows us to express the parameters $u_n^{(k)}$ and
$v_n^{(k)}$ in terms of $\lambda_n^{(k)}$.
We conclude that the dressing functions take the following form
\begin{eqnarray}
D_k(\lambda)=\prod_{n=1}^{M^{(k-1)}}
\frac{\lambda-\lambda_n^{(k-1)}-\frac{\hbar\;(k+1)}{2}}
{\lambda-\lambda_n^{(k-1)}-\frac{\hbar\;(k-1)}{2}}
~~\prod_{n=1}^{M^{(k)}}
\frac{\lambda-\lambda_n^{(k)}-\frac{\hbar\;(k-2)}{2}}
{\lambda-\lambda_n^{(k)}-\frac{\hbar\;k}{2}}\;.
\label{dressingClosed}
\end{eqnarray}

\subsubsection{Universal Bethe equations}

We have chosen the normalisation of the matrix $\wh \cT(\lambda)$ in such a
way that its entries are analytical. Then, the eigenvalues of $\wh
t(\lambda)$ are also analytical, since $\wh t(\lambda)$ can be diagonalised
by a constant matrix (see equation (\ref{comt})).

\begin{theorem}
  The Bethe equations read, for $1 \leq k \leq \enne-1$ and $1 \leq n \leq
  M^{(k)}$
  \begin{eqnarray}
    \label{closedbethe}
    \prod_{m=1}^{M^{(k-1)}}
    e_{-1}\left(\lambda_n^{(k)}-\lambda_m^{(k-1)}\right) \prod_{m=1 \atop
    m\neq n }^{M^{(k)}} e_{2}\left(\lambda_n^{(k)}-\lambda_m^{(k)}\right)
    \prod_{m=1}^{M^{(k+1)}}
    e_{-1}\left(\lambda_n^{(k)}-\lambda_m^{(k+1)}\right) =
    \frac{P_{k}\left(\lambda_n^{(k)}+\frac{\hbar\;k}{2}\right)}
    {P_{k+1}\left(\lambda_n^{(k)}+\frac{\hbar\;k}{2}\right)} \qquad\quad
  \end{eqnarray}
  where
  \begin{eqnarray}
    e_{x}(\lambda) \,=\,
    \frac{\lambda-\frac{\hbar\;x}{2}}{\lambda+\frac{\hbar\;x}{2}}\;.
    \label{ee}
  \end{eqnarray}
  The left hand side of (\ref{closedbethe}) depends only on the choice of
  the algebra (the indices of the function $e_{x}(\lambda)$ describe the
  entries of the Cartan matrix of $gl(\enne)$), while the right hand side
  depends on the choice of the representation.
\end{theorem}
\textbf{Proof:} 
By imposing that the $\Lambda(\lambda)$ residue vanishes at
$\lambda=\lambda_n^{(k)}+\frac{\hbar\;k}{2}$, we find
(\ref{closedbethe}).
\finproof

The RHS of (\ref{closedbethe}) can be written in terms of the functions
$e_{x}(\lambda)$, using the expression of the highest weights. These Bethe
equations have been computed in \cite{ow}, however the method and the
starting hypotheses are different. The identity between the results appears
as a ground for this Ansatz. \\
It should be clear that the Bethe equations (\ref{closedbethe}), and the
dressing of the eigenvalues, (\ref{ClosedEigen}) and
(\ref{dressingClosed}), are valid whatever the expression of the Drinfel'd
polynomial is, and as such are universal. The dressing functions (and thus
the expression of the eigenvalues) appear formally independent from the
choice of the representations. However, the Bethe equations depending on
the representations, their resolution will lead to different eigenvalues.
\\
The choice of a closed spin chain model amounts to the choice of the
$gl(\enne)$ representation $M(\bs{\alpha^k})$ for spins at sites
$k$, $1\leq k \leq\ell$. This will fix the evaluation representations
$M_{\lambda}(\bs{\alpha^k})$, hence the polynomials $P_{k}(\lambda)$. Then,
the eigenvalues and the Bethe equations follow. We now illustrate this
procedure by employing spin chain models.

\bigskip

\textbf{Remark: reducible representations}\\
When the representation is reducible, the above calculations are still
valid, but they do not lead to a complete set of eigenvalues for the
transfer matrix. In fact, one gets in that case all the eigenvalues
associated to the irreducible subrepresentation with highest weight $v^{+}$. A
simple indication for that is the eq. (\ref{qdetv+}) which now implies
(\ref{eq:qdet}) only on this irreducible subrepresentation.

\subsection{Examples \label{sect:ex}}

Choosing appropriate representations, we shall recover known results
associated with the fundamental representation, generalise the relations about
the alternating spin chains and provide new integrable models (such as
general impurity spin chains). For simplicity, we will most of the time set
the inhomogeneous parameters $a_{n}$ to zero. However, our formalism easily
deals with these inhomogeneous parameters, as we shall see in the next
example.

\subsubsection{Closed spin chain in the fundamental
  representation\label{sect:usual}} 

The usual closed spin chain corresponds to spins in the fundamental
representation. The Hamiltonian is given by the well-known formula
\begin{eqnarray}
  \label{Hln}
  H=\left.\frac{d}{d\lambda}\;\big( ln~\wh
  t(\lambda)\big)\right|_{\lambda=0}\,.
\end{eqnarray}
In this case, we have $\bs{\alpha}^n=(1,0,\dots,0)$, for $1\leq n \leq
\ell$. Then, the Drinfel'd polynomials read
\begin{eqnarray}
  P_k(\lambda)=
  \begin{cases}
    \displaystyle
    \prod_{j=1}^\ell(\lambda+a_j-\hbar)&,~~ k=1\\
    \displaystyle
    \prod_{j=1}^\ell(\lambda+a_j)&,~~ k\neq 1
  \end{cases}
\end{eqnarray}
so that 
\begin{eqnarray}
  \frac{P_{k}\left(\lambda_n^{(k)}+\frac{\hbar\;k}{2}\right)}
  {P_{k+1}\left(\lambda_n^{(k)}+\frac{\hbar\;k}{2}\right)} =
  \begin{cases}
    \displaystyle
    \prod_{j=1}^\ell e_1\left(\lambda_n^{(k)}+a_j\right)&,~~ k=1\\
    1&,~~ k\neq 1
  \end{cases}
  \;.
\end{eqnarray}
Plugging these expressions in the Bethe equations (\ref{closedbethe}), we 
recover the usual Bethe equations for closed spin
chains\footnote{We remind that $\hbar=-i$ when dealing with spin chain
models.}
\begin{eqnarray*}
  \prod_{m=1}^{M^{(k-1)}}
  e_{-1}\left(\lambda_n^{(k)}-\lambda_m^{(k-1)}\right) \prod_{m=1 \atop
  m\neq n }^{M^{(k)}} e_{2}\left(\lambda_n^{(k)}-\lambda_m^{(k)}\right)
  \prod_{m=1}^{M^{(k+1)}}
  e_{-1}\left(\lambda_n^{(k)}-\lambda_m^{(k+1)}\right) =
  \begin{cases}
    \displaystyle
    \prod_{j=1}^\ell e_1\left(\lambda_n^{(k)}+a_j\right)&,~~ k=1\\
    1&,~~ k\neq 1
  \end{cases}
\end{eqnarray*}
Since the value of the local operator $\cL_{ij}(\lambda)$ at
$\lambda=0$ is the permutation operator $\cP_{ij}$ between spaces $i$ 
and $j$ (see eq. (\ref{eq:P12})),
we can construct a local Hamiltonian by the relation (\ref{Hln}) (when
$a_{n}=0$)
\begin{eqnarray}
H \propto \sum_{n=1}^\ell \cP_{n-1,n} \qmbox{with} 
\cP_{01}=\cP_{\ell1}\,.
\end{eqnarray}
In the case of $gl(2)$, one recovers the celebrated XXX Hamiltonian
\begin{eqnarray}
H=\frac{1}{2}\sum_{n=1}^\ell 
~\left(~\sigma^x_{n-1}\sigma^x_{n}
+\sigma^y_{n-1}\sigma^y_{n}
+\sigma^z_{n-1}\sigma^z_{n}
+1~\right)\;,
\end{eqnarray}
where $\sigma^t_{n}$, $t=x,y,z$ are the Pauli matrices
at site $n$, and $\sigma^t_{0}=\sigma^t_\ell$.

If we take $a_{p}\neq0$ for a
particular site $p$ (and $a_n=0$ for $n\neq p$), we obtain a Hamiltonian 
with one impurity
\begin{eqnarray*}
H\propto \sum_{n=1 \atop {n\neq p,p+1}}^\ell \cP_{n-1,n} 
-\frac{\hbar}{a_{p}-\hbar}
+\frac{1}{a^2_{p}-\hbar^2}\big(~a_{p}^2\,\cP_{p-1,p+1}-\hbar^2\,\cP_{p+1,p}~\big)
+\frac{\hbar\,a_{p}}{a^2_{p}-\hbar^2}\cP_{p-1,p+1}(\cP_{p-1,p}-\cP_{p+1,p})\;.
\end{eqnarray*}

\subsubsection{Closed spin chain for non-fundamental representations
\label{sect:sspin}}

One can generalise the above example to the case where all the spins 
belong to the same (not necessarily fundamental) representation, 
given by 
\begin{equation}
    \bs{\alpha^1}=\bs{\alpha^2}=\ldots=\bs{\alpha^\ell}=
    (\alpha_{1},\alpha_{2},\dots,\alpha_{\enne})\,.
\end{equation}
In particular, we recover the result given in
\cite{Kul,Tak,Bab,fafa,kit} about the XXX higher spin chains.

We will use the variables
$$
\alpha_{k}\pm\alpha_{k+1}=\beta_{k}^\pm
$$
which are integers, since we consider $gl(\enne)$ irreducible
finite-dimensional representations.
This leads to the following Drinfel'd polynomials
\begin{eqnarray}
  P_k(\lambda)=(\lambda-\hbar\alpha_{k})^\ell \qmbox{so that}
  \frac{P_{k}\left(\lambda_n^{(k)}+\frac{\hbar\;k}{2}\right)}
  {P_{k+1}\left(\lambda_n^{(k)}+\frac{\hbar\;k}{2}\right)} =
  \left[e_{\beta^-_{k}}
  \left(\lambda_n^{(k)}+\hbar\frac{k-\beta_{k}^{+}}{2}\right)\right]^\ell
  \;.
\end{eqnarray}
For instance, if we particularise to the $gl(2)$ spin chain in the
spin $s$ representation, we get as Bethe equations
\begin{eqnarray}
  \label{closed-gl2}
  \prod_{m=1 \atop m\neq n }^{M} e_{2}\left(\lambda_n-\lambda_m\right)
  =\left[e_{2s}\left(\lambda_n+\hbar\frac{1-2s}{2}\right) \right]^\ell\;.
\end{eqnarray}

The construction of a local Hamiltonian cannot be repeated from section
\ref{sect:usual} because there is no particular parameter where the
local operator $\cL(\lambda)$ is the permutation.
However, a local Hamiltonian can be constructed by using the fusion
method introduced in \cite{Kul,Tak,Bab,kit} or by evaluating the
universal $R$-matrix, see e.g. \cite{fafa}. It takes the form
\begin{eqnarray}
  H\propto \sum_{n=1}^{\ell} Q_{2s} \left(~s^x_{n-1}s^x_{n}
  +s^y_{n-1}s^y_{n} +s^z_{n-1}s^z_{n} ~\right)\,,\qmbox{where}
  Q_{2s}(x) = \sum_{j=1}^{2s} \left(\sum_{k=1}^j \frac{1}{k}\right)\,
  \prod_{l=0\atop l\neq j}^{2s} \frac{x-x_{l}}{x_{j}-x_{l}}\,.
\end{eqnarray}
In the above formula, $x_{l}=\frac{1}{2}(l(l+1)-2s(s+1))$, $s^t_{n}$, 
$t=x,y,z$ are the $gl(2)$ generators  in the spin $s$ representation acting 
in the quantum space $n$, and satisfying
$s^t_{0}=s^t_{\ell}$. The energy spectrum is then given by
\begin{eqnarray}
E=-\sum_{j=1}^\ell \frac{s}{\lambda_j+s^2}\;,
\end{eqnarray}
where $\lambda_j$ are solutions of the Bethe equations (\ref{closed-gl2}).

\subsubsection{Alternating spin chains\label{sect:aspin}}

In alternating spin chains, the spins along the chain belong alternatively to 
two different given representations. We take the
particular example of the alternating spin chain with
the number of sites $\ell=2\wt \ell$ even. The spins of even sites are
represented in the fundamental representation whereas the spins of the
odd sites are in another representation. We take the following
particular example where the highest weights are given by
\begin{eqnarray}
\bs{\alpha}^n=
\begin{cases}
(1,0,\dots,0),&~~ 1\leq n \leq \ell \qmbox{and} n~\mbox{even}\\
(2,0,\dots,0),&~~ 1\leq n \leq \ell \qmbox{and} n~\mbox{odd}
\end{cases}
~.
\end{eqnarray}
Then, the left hand side of the Bethe equations read, for $1\leq n\leq 
M^{(k)}$, $1\leq k\leq \enne-1$ and $\enne>2$
\begin{eqnarray}
\frac{P_{k}\left(\lambda_n^{(k)}+\frac{\hbar\;k}{2}\right)}
{P_{k+1}\left(\lambda_n^{(k)}+\frac{\hbar\;k}{2}\right)}
=
\begin{cases}
\left(
e_1\left(\lambda_n^{(1)}\right)
e_2\left(\lambda_n^{(1)}-\frac{\hbar}{2}\right)
\right)^{\wt \ell} &,~~ k=1\\
1&,~~ 1<k<\enne
\end{cases}
\;.
\end{eqnarray}
We recover the Bethe equation given in \cite{dewo} (see also
\cite{mnanal2}) for $gl(2)$. In the case $gl(2)$, we can compute a
Hamiltonian by the usual formula (\ref{Hln}) which contains both
nearest and next-to-nearest neighbour interactions with periodic
boundary conditions. The explicit form of this Hamiltonian is given by
\cite{dewo}
\begin{eqnarray}
H&\propto&\sum_{j=1}^{\wt \ell}
\{~
(2\bs{\sigma}_{2j}\cdot \bs{s}_{2j+1}+1)
(2\bs{\sigma}_{2j+2}\cdot \bs{s}_{2j+1}+3)
\nonumber\\
&&\hspace{1cm}+
(2\bs{\sigma}_{2j}\cdot \bs{s}_{2j-1}+1)
\left[
(\bs{s}_{2j-1}\cdot \bs{s}_{2j+1}+1)
(2\bs{\sigma}_{2j}\cdot \bs{s}_{2j+1}+1)
+2
\right]~
\}
\end{eqnarray}
where $\bs{\sigma}=(\sigma^x,\sigma^y,\sigma^z)$ are the Pauli
matrices and $\bs{s}=(s^x,s^y,s^z)$ are the generators of $sl(2)$ in
the spin 1 representation. 

We can also recover the results of \cite{abad2} where another type of
alternating spin chains has been studied for $su(3)$.

\subsubsection{Impurity}

We consider now a spin chain with one site (the impurity) in a
representation different from the others. Let us take as example a spin
chain where all sites are represented in the fundamental representation
except for the $p^{th}$ which is associated to the representation of
highest weight ${\bs \alpha^p}$. In this case, the left hand side of
(\ref{closedbethe}) becomes
\begin{eqnarray}
  \frac{P_{k}\left(\lambda_n^{(k)}+\frac{\hbar\;k}{2}\right)}
  {P_{k+1}\left(\lambda_n^{(k)}+\frac{\hbar\;k}{2}\right)} =
  \begin{cases}
    \displaystyle e_1\left(\lambda_n^{(k)}\right)^{\ell-1}
    \frac{\lambda_n^{(k)}+\frac{\hbar}{2}-\hbar\;\alpha_1^p}
    {\lambda_n^{(k)}+\frac{\hbar}{2}-\hbar\;\alpha_2^p} &,~~ k=1\\
    & \\
    \displaystyle
    \frac{\lambda_n^{(k)}+\frac{\hbar\;k}{2}-\hbar\;\alpha_k^p}
    {\lambda_n^{(k)}+\frac{\hbar\;k}{2}-\hbar\;\alpha_{k+1}^p} 
    &,~~ k\neq 1
  \end{cases}
\end{eqnarray}
The Hamiltonian can be written as
\begin{eqnarray}
  H \propto \sum_{n=1 \atop {n\neq p,p+1}}^\ell \cP_{n-1,n}
  +~\big(~\cP_{p-1,p+1}~\wh\cL_{p+1,p}(0)-\hbar~\big)~\wh\cL_{p-1,p}^{-1}(0)
\end{eqnarray}
where $\cP_{01}=\cP_{\ell1}$, $\wh\cL_{n,p}(\lambda)=\sum_{i,j}
E_{ij}\otimes (\lambda-\hbar \cE_{ji})$. In the last formula, $E_{ij}$
belongs to the space $n$ (fundamental representation), while
$(\lambda-\hbar \cE_{ji})$ is in the particular space $p$ where the
generators of $gl(\enne)$, $\cE_{ij}$, are in the representation with the
highest weight ${\bs \alpha^p}$.

\subsubsection{Generalisation to tensor products of representation 
on each site}

Up to now, we have assumed that on each site of the spin chain, only 
one evaluation representation occurs. This assumption is natural from 
the spin chain point of view, since one can interpret the underlying 
$gl(\enne)$ representation as carrying the spin. However, the 
algebraic framework we have presented can deal with more general 
representations of the Yangians, provided they are irreducible and 
finite-dimensional. Using the theorem \ref{the:irr}, the irreducible 
representations will be constructed from tensor products of evaluation 
representations. Let us stress that, generically, this tensor product 
of evaluation representations {is} \underline{irreducible}, although for 
the transfer matrix symmetry algebra $gl(\enne)$ these representations are 
fully reducible.  

{}From the physical point of view, the model will describe a spin chain
possessing on each site a quantum space which is a tensor product of
evaluation representations of the Yangian. However, this model (in
particular the transfer matrix) can be reinterpreted as a usual spin chain
model but with a higher number of sites, each of them associated to only
one evaluation representation.

Finally, let us remark that this construction is in essence opposite to the
fusion procedure. Indeed, for the fusion, one takes particular points
(described in theorem \ref{the:irr}) where the tensor product of evaluation
representations is \underline{reducible}.

\section{Open spin chains with preserving boundary conditions}

In this section, we compute, along the lines described in the previous
section,  
the Bethe equations for the open spin chains with soliton preserving boundary
conditions \cite{dvgr,GZ,abad,done}. For such a purpose, we first need to
introduce some new algebraic objects such as the reflection algebra or the
$K$ matrix.

\subsection{Reflection $K$ matrix}

In the case of soliton preserving boundary conditions, we need to introduce
numerical matrices, called $K$ matrices, which are solutions of the
reflection (boundary Yang--Baxter) equations \cite{cherednik}:
\begin{eqnarray}
  R_{ab}(\lambda_{a}-\lambda_{b})\ K_{a}(\lambda_{a})\
  R_{ba}(\lambda_{a}+\lambda_{b})\ K_{b}(\lambda_{b})=
  K_{b}(\lambda_{b})\
  R_{ab}(\lambda_{a}+\lambda_{b})\ K_{a}(\lambda_{a})\
  R_{ba}(\lambda_{a}-\lambda_{b}) \;.\qquad
  \label{re}
\end{eqnarray}
The $K$ matrix is interpreted as the reflection of a
soliton on the boundary, coming back as a soliton.\\
The solutions of the equation (\ref{re}) have been classified in
\cite{Mint}:
\begin{proposition}
  \label{prop:SP1}
  Any invertible solution of the soliton preserving 
  reflection equation (\ref{re}) takes the form $\displaystyle
  K(\lambda) =
   U\,\left(\,\EE\,+\,\frac{\xi}{\lambda}\,\II_\enne\,\right)U^{-1}$ 
  where either
  \begin{itemize}
  \item[(i)]
    $\EE$ is diagonal and $\EE^2=\II_\enne$ (diagonalisable solutions)
  \item[(ii)]
    $\EE$ is strictly triangular and $\EE^2=0$
    (non-diagonalisable solutions)
  \end{itemize}
  The matrix $U$ is an element of the group $GL({\enne})$ and $\xi$ a
  free parameter. The classification is done up to multiplication by a
  function of the spectral parameter.\\
  Note that all the $K$ matrices (but zero)
  obey a relation
  $K(\lambda)K(-\lambda)=f(\lambda)\,\II_{\enne}$ for some
  non-zero even
  function $f$.
\end{proposition}

A suitable relabelling of the indices allows us to choose the matrix
$\EE$ in $(i)$ of proposition \ref{prop:SP1} as 
\begin{eqnarray}
\EE=diag(\underbrace{1,\ldots,1}_{\emme},
\underbrace{-1,\ldots,-1}_{\enne-\emme})\;,
\end{eqnarray}
with $0\leq \emme \leq \enne$. 
In the following we will only deal with diagonal solutions of the form
\begin{eqnarray}
\wh K(\lambda)=diag(\underbrace{\lambda+\xi,\ldots,\lambda+\xi}_{\emme}~,~
\underbrace{-\lambda+\xi,\ldots,-\lambda+\xi}_{\enne-\emme})\;.
\end{eqnarray}
We normalise the $K$ matrix so that its entries be analytical.\\

\paragraph{Case of non-diagonal reflection matrices: }
The general treatment of non-diagonal reflection matrices is yet an open
problem. In the case where each spin is represented in the fundamental
representation, the problem has been solved in \cite{lepetit,selene}
for $K^+=1$ and in \cite{gama} for simultaneously diagonalisable
reflection matrices $K^+$ and $K^-$. In the
case of the $XXZ$ model, the procedure given in \cite{lepetit,selene} to
treat the non-diagonal reflection matrices does not work. However,
interesting developments have been done in \cite{nepoND,chinois} to
attempt a general treatment of non-diagonal reflection matrices.\\

\subsection{Reflection algebra}

The reflection algebras are constructed as subalgebras of a Yangian, which 
is here $\yy(gl(\enne))$. Starting from the generators $\cT(\lambda)$
of $\yy(gl(\enne))$ introduced in (\ref{def:T}), we define
\begin{equation}
  \label{eq:defB}
  \cB(\lambda)=\cT(\lambda)\,K(\lambda) \,\cT(-\lambda)^{-1}\;. 
\end{equation}
$\cB(\lambda)$ generates an algebra, denoted $\fB(\enne,\emme)$, whose
exchange 
relations are given by
\begin{eqnarray}
&& R_{ab}(\lambda_{a}-\lambda_{b})\ \cB_{a}(\lambda_{a})\
  R_{ba}(\lambda_{a}+\lambda_{b})\ \cB_{b}(\lambda_{b})=
  \cB_{b}(\lambda_{b})\
  R_{ab}(\lambda_{a}+\lambda_{b})\ \cB_{a}(\lambda_{a})\
  R_{ba}(\lambda_{a}-\lambda_{b}) \;.\qquad
  \label{re-algebra}
 \end{eqnarray}
Writing $\cB(\lambda)$ as
$$
\cB(\lambda)=\sum_{i,j=1}^\enne E_{ij}\otimes B_{ij}(\lambda)
=\sum_{n=0}^{+\infty} \frac{\cB^{(n)}}{\lambda^{n}}
$$
one can show that $\cB^{(1)}$ generates a $gl(\emme)\oplus gl(\enne-\emme)$
subalgebra in $\fB(\enne,\emme)$.\\
Another reflection matrix $K^+(\lambda)$, solution of an equation dual
to (\ref{re}), is usually introduced to study open spin chains
\cite{sklyanin}. For simplicity, 
we will take here $K^+(\lambda)=\II_\enne$.\\

The (algebraic) monodromy matrix used to construct open spin chain is
obtained from the local operators $\cL_{a\,j}(\lambda)$ of the Yangian
(\ref{mono}). It takes the following form
\begin{eqnarray}
\cB_a(\lambda)=
\cL_{a1}(\lambda)\;\dots\;\cL_{a\ell}(\lambda)~
K_a(\lambda)~
\cL_{a\ell}^{-1}(-\lambda)\;\dots\;\cL_{a1}^{-1}(-\lambda)\;.
\label{monoRE}
\end{eqnarray}
The transfer matrix becomes
\begin{eqnarray}
b(\lambda)=tr_a\left(\cB_a(\lambda)\right)=
\sum_{i=1}^\enne B_{ii}(\lambda)
\end{eqnarray}
and, as in the Yangian case, the commutation relations defining
the algebra allow us to show
\begin{eqnarray}
\label{comtmre}
[b(\lambda),b(\mu)]=0\;.
\end{eqnarray}
This relation (\ref{comtmre}) guarantees the integrability of the model,
usually described by the following Hamiltonian
\begin{eqnarray}
   H = -\frac{1}{2} \frac{d}{d \lambda}b(\lambda)\Big\vert_{\lambda =0}\;.
\end{eqnarray}
Anticipating again the physical spin chain interpretation, one can compute the 
symmetry of these models
\begin{proposition}
    The transfer matrix $b(\lambda)$ describing  
    open spin chain models admits an $gl(\emme)\oplus gl(\enne-\emme)$ 
    symmetry.
\end{proposition}
\textbf{Proof:} Following the steps given for the closed spin chains 
(see proof of proposition \ref{prop:symt}), 
one shows that $[\cB^{(1)}\,,\,b(\lambda)]=0$. Since $\cB^{(1)}$ 
generates a $gl(\emme)\oplus gl(\enne-\emme)$  algebra, this ends the 
proof.
\finproof\\

\subsection{Representations of $\fB(\enne,\emme)$}

\subsubsection{Representation of $\cT^{-1}(\lambda)$}

In order to study the representations of $\fB(\enne,\emme)$, we start from
the representations of the Yangian introduced in the section \ref{reps}.
Let $M_\lambda({\bs \alpha})$ be an evaluation representation of
$\cL(\lambda)$ with the highest weight vector $v$. We can show that $v$ is
also a highest vector of $\cL^{-1}(\lambda)$ with
\begin{eqnarray}
  \label{HWinv1}
  &&L'_{jk}(\lambda)\;v=0 \qmbox{,} 1\leq k<j \leq \enne \\
  \label{HWinv2}
  &&L'_{kk}(\lambda)\;v=\lambda\,
  \frac{
  \left(\lambda+\hbar-\hbar\;\alpha_1 \right)
  \cdots
  \left(\lambda+k\hbar-\hbar-\hbar\;\alpha_{k-1}\right)
  }{
  \left(\lambda-\hbar\;\alpha_1\right)
  \cdots
  \left(\lambda+k\hbar-\hbar-\hbar\;\alpha_k\right)
  }\; v 
  \qmbox{,}  1\leq k \leq \enne \;,
\end{eqnarray}
where $L'_{jk}(\lambda)$ are the matrix elements of $\cL^{-1}(\lambda)$.
The values appearing in (\ref{HWinv2}) are computed from the relation
(\ref{comp:inv}).\\
The relations (\ref{HWinv1}) and (\ref{HWinv2}) imply that $v^+$ as given
in (\ref{v+}) is the highest weight vector of $\cT^{-1}(\lambda)$ with
\begin{eqnarray}
  \label{HWmonoinv1}
  &&T'_{jk}(\lambda)\;v^+=0 \qmbox{,} 1\leq k<j \leq \enne  \\
  \label{HWmonoinv2}
  &&T'_{kk}(\lambda)\;v^+=\lambda^\ell
  \frac{
  P_{1}(\lambda+\hbar)
  \cdots
  P_{k-1}(\lambda+k\hbar-\hbar)
  }{
  P_{1}(\lambda)
  \cdots
  P_{k}(\lambda+k\hbar-\hbar)
  }\; v^+ 
  \qmbox{,}  1\leq k \leq \enne \;,
\end{eqnarray}
where $T'_{jk}(\lambda)$ are the matrix elements of the matrix
$\cT^{-1}(\lambda)$ and $P_{k}(\lambda)$ are defined in 
(\ref{drinP}).\\
Let us remark that 
\begin{equation}
 \widetilde\cT(\lambda)= \frac{P_{1}(-\lambda)\,P_{2}(-\lambda+\hbar)\cdots
 P_{\enne}(-\lambda+\hbar\enne-\hbar)}{(-\lambda)^{\enne\ell}}\,\cT^{-1}(-\lambda)
 \label{Ttilde}
\end{equation}
can be understood as the Yangian generators represented in the following
tensor product of evaluation representations (as defined in section
\ref{decadix})
\begin{eqnarray}
  \bigotimes_{n=1}^\ell\Big(M_{\lambda}(\bs{\beta^{1,n}})\otimes\cdots\otimes
  M_{\lambda}(\bs{\beta^{\enne,n}}) \Big)\label{evalTinv}
\end{eqnarray}
where
$\bs{\beta^{k,n}}=(1-\alpha^n_1,\dots,k-1-\alpha^n_{k-1},0,k-\alpha^n_{k+1},
\dots,\enne-1-\alpha^n_\enne)$.
This shows that the matrix 
$(-\lambda)^{(\enne-1)\ell}\,\widetilde\cT(\lambda)$ is 
analytical.

\subsubsection{Representation of the monodromy matrix $\cB(\lambda)$}

We can now describe the representations of the monodromy matrix
$\cB(\lambda)$ defined by (\ref{monoRE}).
It is known \cite{momo} that any finite-dimensional representation of
$\fB(\enne,\emme)$ is a highest weight representation. They can be 
constructed in the following way
\begin{theorem}
    Let us consider the Yangian highest weight representation 
    $M_{\lambda}(\bs{\alpha^1})\otimes 
    \ldots\otimes M_{\lambda}(\bs{\alpha^\ell})$ with highest weight 
    vector $v^+=v^1\otimes\ldots\otimes v^\ell$. Then the realisation
   (\ref{monoRE}) generates a $\fB(\enne,\emme)$ 
    highest weight representation, whose highest weight vector is 
    also $v^+$ with
\begin{eqnarray}
\label{HWB1}
&&B_{jk}(\lambda)\;v^+=0 \,,\qquad 1\leq k<j \leq \enne  \\
\label{HWB2}
&&B_{kk}(\lambda)\;v^+=\left(
\sum_{j=1}^{k-1}
a_j(\lambda)~{\mu}_j(\lambda)
+\frac{2\lambda}{2\lambda-k\hbar+\hbar}\,{\mu}_k(\lambda)\right)\,v^+
\,,\qquad 1\leq k \leq \enne\;,\qquad
\end{eqnarray}
where, for $1\leq j \leq \enne$,
\begin{eqnarray}
a_j(\lambda)=\frac{-2\lambda}{(2\lambda-j\hbar+\hbar)(2\lambda-j\hbar)}
\end{eqnarray}
and 
\begin{eqnarray}
{\mu}_k(\lambda)=
\begin{cases}
\displaystyle
(-1)^\ell
(\lambda+\xi)
P_k(\lambda)
\frac{P_{1}(-\lambda+\hbar)\cdots
P_{k-1}(-\lambda+k\hbar-\hbar)
}{P_{1}(-\lambda)\cdots
P_{k}(-\lambda+k\hbar-\hbar)
}\qmbox{for}1\leq k \leq \emme\\
&\\
\displaystyle
(-1)^\ell
(-\lambda+\xi+\emme\hbar)
P_k(\lambda)
\frac{P_{1}(-\lambda+\hbar)\cdots
P_{k-1}(-\lambda+k\hbar-\hbar)
}{P_{1}(-\lambda)\cdots
P_{k}(-\lambda+k\hbar-\hbar)
}\qquad \emme+1\leq k \leq \enne\\
\end{cases}
\end{eqnarray}
\end{theorem}
\textbf{Proof:} 
A direct calculation (similar to the one done in \cite{momo}) leads
to, for $1\leq k \leq \emme$,
\begin{eqnarray}
\label{relkM}
\frac{2\lambda-k\hbar+\hbar}{2\lambda}B_{kk}(\lambda)v^++
\frac{\hbar}{2\lambda}\sum_{j=1}^{k-1}B_{jj}(\lambda)v^+
=(\lambda+\xi)~T_{kk}(\lambda)T'_{kk}(-\lambda)v^+
\end{eqnarray}
and for $\emme+1 \leq k \leq \enne$
\begin{eqnarray}
\label{relkN}
\frac{2\lambda-k\hbar+\hbar}{2\lambda}B_{kk}(\lambda)v^+
+
\frac{\hbar}{2\lambda}
\sum_{j=1}^{k-1}B_{jj}(\lambda)v^+
=(-\lambda+\xi+\emme\hbar)~T_{kk}(\lambda)T'_{kk}(-\lambda)v^+\;.
\end{eqnarray}
Then, inverting these formulae and using the expressions 
(\ref{HWmono2}) and (\ref{HWmonoinv2}), one
gets the expression (\ref{HWB2}).
\finproof

\subsection{Analytical Bethe Ansatz}

The analytical Bethe Ansatz method is based upon the analyticity of
the represented generators of the algebra. It is ensured by a suitable
normalisation given in the following proposition
\begin{proposition}
    Let 
    \begin{eqnarray}
\wh\cB(\lambda)=(-1)^\ell~
P_{1}(-\lambda)\cdots P_{\enne}(-\lambda+\enne\hbar-\hbar)~
\cB(\lambda)\,.
\end{eqnarray}
Then, $\wh\cB(\lambda)$ is analytical (in $\lambda$).
\end{proposition}
\textbf{Proof:} 
$\wh\cB(\lambda)$ can be rewritten as 
$$
\wh\cB(\lambda)=\wh\cT(\lambda)\,\times\,\wh K(\lambda)
\,\times\,\left((-\lambda)^{(\enne-1)\ell}\,\widetilde\cT(\lambda)\right)\,.
$$
The three terms of this product are analytical.
\finproof\\
{}From now on, we will use $\widehat{\cB}(\lambda)$ instead of $\cB(\lambda)$
to ensure, as in the closed spin chain case, the analyticity of the
eigenvalues of the transfer matrix
\begin{equation}
    \widehat b(\lambda)=tr_{a}\,\wh\cB_{a}(\lambda)\,.
\end{equation}

\subsubsection{Pseudo-vacuum}

As in the case of the closed spin chain, the first step of the
analytical Bethe Ansatz consists in finding a particular eigenvalue of
the transfer matrix. This eigenvalue is computed thanks to the highest
weight vector $v^+$. Indeed, one gets
\begin{eqnarray}
\wh b(\lambda)\;v^+=\sum_{k=1}^\enne \wh B_{kk}(\lambda)v^+
=\Lambda^0(\lambda)\;v^+
\end{eqnarray}
where
\begin{eqnarray}
\Lambda^0(\lambda)=\sum_{k=1}^\enne g_k(\lambda)~\beta_k(\lambda)\;.
\end{eqnarray}
The functions $g_k(\lambda)$ depends only on the boundary matrix. They
are given by
\begin{eqnarray}
g_k(\lambda)=
\frac{2\lambda(2\lambda-\enne\hbar)}
{(2\lambda-k\hbar+\hbar)(2\lambda-k\hbar)}
\times
\begin{cases}
\lambda+\xi\qmbox{for}1\leq k \leq \emme\\[1.2ex]
-\lambda+\xi+\emme\hbar\qmbox{for}\emme+1\leq k \leq \enne\,.
\end{cases}
\label{fg}
\end{eqnarray}
The functions $\beta_k(\lambda)$ depend on the choice of the
representation:
\begin{eqnarray}
\beta_k(\lambda)=
P_{1}(-\lambda+\hbar)
\cdots
P_{k-1}(-\lambda+k\hbar-\hbar)
P_k(\lambda)
P_{k+1}(-\lambda+k\hbar)
\cdots
P_{\enne}(-\lambda+\enne\hbar-\hbar)\;.
\label{fb}
\end{eqnarray}
Let us remark that $\Lambda^0(\lambda)$ is analytical and in particular
its residue for $\lambda=k\hbar/2$ vanishes ($0\leq k \leq \enne$).

\subsubsection{Dressing functions}

The central hypothesis of the analytical Bethe Ansatz is that
all the eigenvalues of $\widehat{b}(\lambda)$ can be written
\begin{eqnarray}
\Lambda(\lambda)=\sum_{k=1}^\enne
g_k(\lambda)~\beta_k(\lambda)~D_k(\lambda)\;,
\end{eqnarray}
where the dressing functions $D_k(\lambda)$ are rational functions and need
to be determined while $g_k(\lambda)$ and $\beta_k(\lambda)$ are given by
(\ref{fg}) and (\ref{fb}), respectively. The vanishing of the residues of
$\Lambda(\lambda)$ at $\lambda=k\hbar/2$ implies that
\begin{eqnarray}
D_{k}(k\hbar/2)=D_{k+1}(k\hbar/2)\qmbox{for}1\leq k \leq \enne-1\;.
\end{eqnarray}
Starting from the expression (\ref{dress:funct}) for the dressing
functions, one can show that the $M^{(k)}$'s are even, and that (up to a
rescaling $M^{(k)}\to M^{(k)}/2$) the dressing functions read
\begin{eqnarray}
D_{k}(\lambda)&=&
\prod_{n=1}^{M^{(k-1)}}
\frac{\lambda+\lambda_n^{(k-1)}-\frac{\hbar(k+1)}{2}}
{\lambda+\lambda_n^{(k-1)}-\frac{\hbar\;(k-1)}{2}}
\quad
\frac{\lambda-\lambda_n^{(k-1)}-\frac{\hbar(k+1)}{2}}
{\lambda-\lambda_n^{(k-1)}-\frac{\hbar\;(k-1)}{2}}
\nonumber\\
&&\times
\prod_{n=1}^{M^{(k)}}
\frac{\lambda+\lambda_n^{(k)}-\frac{\hbar k}{2}+\hbar}
{\lambda+\lambda_n^{(k)}-\frac{\hbar\;k}{2}}
\quad
\frac{\lambda-\lambda_n^{(k)}-\frac{\hbar k}{2}+\hbar}
{\lambda-\lambda_n^{(k)}-\frac{\hbar\;k}{2}}\;,
\end{eqnarray}
where $M^{(0)}=M^{(\enne)}=0$.

\subsubsection{Bethe equations}

The normalisation of the matrix $\wh \cB(\lambda)$ has been chosen in
such a way that its entries are analytical. Then, the eigenvalues of
the transfer matrix $b(\lambda)$ are also analytical (since the
diagonalisation matrix does not depend on $\lambda$). 
\begin{theorem}
The Bethe equations read, for $1 \leq k \leq \enne-1$ and
$1 \leq n \leq M^{(k)}$ 
\begin{eqnarray}
\label{REbethe}
&&\hspace{-2cm}
\prod_{m=1}^{M^{(k-1)}}
\wt{e}_{-1}\left(\lambda_n^{(k)},\lambda_m^{(k-1)}\right)
\prod_{m=1 \atop m\neq n}^{M^{(k)}}
\wt{e}_{2}\left(\lambda_n^{(k)},\lambda_m^{(k)}\right)
\prod_{m=1}^{M^{(k+1)}}
\wt{e}_{-1}\left(\lambda_n^{(k)},\lambda_m^{(k+1)}\right)
\nonumber\\
&=&
\frac{\beta_{k}\left(\lambda_n^{(k)}+\frac{\hbar\;k}{2}\right)}
{\beta_{k+1}\left(\lambda_n^{(k)}+\frac{\hbar\;k}{2}\right)}
\times
\begin{cases}
-e_{-\emme-2\xi/\hbar}\left(\lambda_m^{(\emme)}\right)\qmbox{if}k=\emme
\vspace{3mm}\\
1 \qmbox{otherwise} 
\end{cases}
\end{eqnarray}
where 
\begin{eqnarray}
\wt{e}_{x}(\lambda,\mu)=e_{x}\left(\lambda-\mu\right)
e_{x}\left(\lambda+\mu\right)\;,
\end{eqnarray}
the functions $e_{x}\left(\lambda\right)$ are defined by
(\ref{ee}) and $M^{(0)}=M^{(\enne)}=0$.
\end{theorem}
\textbf{Proof:} 
By imposing the vanishing of the $\Lambda(\lambda)$ residue at
$\lambda=\lambda_n^{(k)}+\frac{\hbar\;k}{2}$, we obtain (\ref{REbethe}).
\finproof\\
As in the case of the Yangian, the left hand side of (\ref{REbethe})
depends only on the choice of the algebra whereas the right hand side depends
on the choice of the representation and the $K$ matrix. 

\subsection{Examples}

\subsubsection{Generalities}

All the cases presented in section \ref{sect:ex} can be treated in a
similar way 
for the open spin chain, using the usual formula given in \cite{sklyanin}
for the Hamiltonian.

As a basic example, one can easily check that the present approach
reproduces correctly the results obtained for the open $gl(\enne)$-spin
chain with generic boundary \cite{selene}.

As more involved examples, we can generalise directly the spin $s$
chain and the 
alternating spin chain (see sections \ref{sect:sspin} and
\ref{sect:aspin}) by adding a boundary with the procedure given above, 
extending the results obtained in
\cite{ana}. 

\subsubsection{Boundaries with operators}

One may wonder whether the boundary matrices $K_{\pm}$ can be promoted 
to operators. Indeed this amounts to ``fuse'' the boundary to the last 
site to get a dynamical boundary. This was considered for instance in
\cite{sklyanin,zab,lf,doma,pasc}.

We treat here an example suggested by K. Zarembo. We study the $gl(2)$ spin
chain with $\ell-2$ spins $1$ in the bulk and for two spins $1/2$ the
boundaries. For this spin chain, we represent the monodromy matrix where
the highest weights are given by
\begin{eqnarray}
\bs{\alpha^n}=
\begin{cases}
(1,0)&\qmbox{,}n=1,\ell\\
(2,0)&\qmbox{,}1<n<\ell
\end{cases}
\end{eqnarray}

The corresponding integrable Hamiltonian is given by (up to an
irrelevant overall normalisation)
\begin{eqnarray}
  \label{eq:HZarembo}
  H = 
     2 {\bs\sigma}_1 \cdot {\bs S}_2  
    + 2 {\bs S}_{L-1}  \cdot {\bs\sigma}_L +
    \sum_{i=2}^{\ell-2} \left({\bs S}_i \cdot {\bs S}_{i+1}
    - \frac14 \left( {\bs S}_i \cdot {\bs S}_{i+1} \right)^2
  \right) 
\end{eqnarray}
with the following conventions:
\begin{equation}
  \label{eq:sigma}
  \sigma^+ = \begin{pmatrix} 0 & 1 \\ 0 & 0 \end{pmatrix}\;, \qquad
  \sigma^- = \begin{pmatrix} 0 & 0 \\ 1 & 0 \end{pmatrix}\;, \qquad
  \sigma^z = \begin{pmatrix} 1 & 0 \\ 0 & -1 \end{pmatrix} \;,
\end{equation}

\begin{equation}
  \label{eq:Smatrix}
  S^+ = \sqrt2 \begin{pmatrix} 0 & 1 & 0 \\ 0 & 0 & 1 \\ 0 & 0 & 0 
  \end{pmatrix}\;, \qquad
  S^- = \sqrt2 \begin{pmatrix} 0 & 0 & 0 \\ 1 & 0 & 0 \\ 0 & 1 & 0 
  \end{pmatrix}\;, \qquad
  S^z = \begin{pmatrix} 2 & 0 & 0 \\ 0 & 0 & 0 \\ 0 & 0 & -2 
  \end{pmatrix}\;, \qquad
\end{equation}
and
\begin{equation}
  \label{eq:scalprod}
  {\bs A} \cdot {\bs B} = 2 A^+ B^- + 2 A^- B^+ +  A^z B^z \;.
\end{equation}
The $\bs S$ used here is actually equal to $2\bs s$ of section \ref{sect:ex}
so that $\bs\sigma $ and $\bs S$ have the same commutation relations.

This Hamiltonian comes from the monodromy matrix (\ref{eq:defB}) with
the following prescriptions: the auxiliary space is three-dimensional
(spin 1 representation of $gl(2)$); the reflection matrices $K^\pm$ are
taken to be identity matrices.
\\
Hence $\tilde b(\lambda) =  \tr_a \cT_a(\lambda) \cT_a(-\lambda)^{-1}$
with 
\begin{equation}
  \label{eq:TZarembo}
  \cT(\lambda) = R_{a\ell}^{(1,\frac12)}(\lambda)
  R_{a,\ell-1}^{(1,1)}(\lambda)  \cdots R_{a2}^{(1,1)}(\lambda) 
  R_{a1}^{(1,\frac12)}(\lambda) 
\end{equation}
and
\begin{eqnarray}
  \label{eq:R3o2}
  R_{aj}^{(1,\frac12)}(\lambda) &=& \lambda \II_3 \otimes \II_2
  - \frac\hbar2 {\bs S}_a \cdot {\bs \sigma}_j \qquad j=1,\ell \;,
  \\[2mm]
  \label{eq:R3o3}
  R_{aj}^{(1,1)}(\lambda) &=& 
  \frac{(\lambda+\hbar)(\lambda-2\hbar)}{2\hbar^2} \II_3 \otimes \II_3
  - \frac{\lambda-\hbar}{4\hbar}   {\bs S}_a \cdot \bs S_j 
  + \frac1{16} \left(\bs S_a \cdot \bs S_j \right)^2\;,
  \quad j=2,\dots, \ell-1 \;. \quad 
\end{eqnarray}

These $R$-matrices can be derived from 
$R_{ij}^{(\frac12,\frac12)}$ given by (\ref{r}) using the usual fusion
procedure \cite{mnf}. 
As a consequence, the transfer matrix $\tilde b(\lambda)$ (and the
Hamiltonian) commutes with the transfer matrix $b(\lambda)$ built with
the same quantum spaces and spin $1/2$ auxiliary space.

Then, using (\ref{REbethe}), the
Bethe equations are, for $1\leq n \leq M$
\begin{eqnarray}
\label{REbethe-ex}
\prod_{m=1 \atop m\neq n}^{M}
e_{2}\left(\lambda_n-\lambda_m\right)
e_{2}\left(\lambda_n+\lambda_m\right)
=
e_{1}\left(\lambda_n\right)^{\ell+2}
e_{3}\left(\lambda_n\right)^{\ell-2}
\end{eqnarray}

The bulk part of (\ref{eq:HZarembo}) is the mixing matrix for some 
sort of gluon operators in large-$N$ QCD, see \cite{Zarembo}. 
The spin chain boundary term in (\ref{eq:HZarembo}) 
corresponds to the  quark-gluon operators.

\subsection*{Perspectives}
A natural development of this work is the generalisation to the
trigonometric case, which will be presented in a further
publication. Soliton non-preserving boundary conditions will also be
studied in this framework.

\vskip0.5cm

\textbf{Acknowledgements:} This work is supported by the TMR Network 
`EUCLID. Integrable models and applications: from strings to condensed
matter', contract number HPRN-CT-2002-00325.\\
We would like to thank K. Zarembo for suggesting the last example.

\end{document}